\documentclass[journal,twoside,web]{IEEEtran}

\usepackage{amssymb}

\usepackage[figuresright]{rotating}
\usepackage{subcaption}
\usepackage{amsmath}
\usepackage[inline]{enumitem}
\usepackage{xcolor}

\usepackage{caption}
\usepackage{subcaption}

\usepackage{verbatim}

\begin{document}

\title{
	{\fontsize{18.6}{23}\selectfont Acoustical Features as Knee Health Biomarkers: A Critical Analysis}
}

\author{
\IEEEauthorblockN {
   Christodoulos Kechris$^{a,*}$,
   Jerome Thevenot$^{a}$,     
   Tomas Teijeiro$^{a,b}$,
   Vincent A. Stadelmann$^{c}$,
   Nicola A. Maffiuletti$^{d}$,
   David Atienza$^{a}$
}

\IEEEauthorblockA {
  $^{a}$\textit{\small Embedded Systems Laboratory (ESL), Ecole Polytechnique Federale de Lausanne (EPFL), Switzerland} \\
  $^{b}$\textit{\small Basque Center for Applied Mathematics (BCAM), Spain} \\
  $^{c}$\textit{\small Department of Research and Development, Schulthess Klinik, Zürich, Switzerland} \\
  $^{d}$\textit{\small Human Performance Lab, Schulthess Klinik, Zürich, Switzerland}
}
\thanks{Funding: This work was supported by the Swiss Innovation Agency (grant Innosuisse 59444.1 IP-LS) and Wilhelm-Schulthess Stiftung. T. Teijeiro is supported by the grant RYC2021-032853-I funded by MCIN/AEI/ 10.13039/501100011033 and by European Union NextGenerationEU/PRTR.}
\thanks{*Corresponding author C.K. e-mail: christodoulos.kechris@epfl.ch}
}

\maketitle

\begin{abstract}
Acoustical knee health assessment has long promised an alternative to clinically available medical imaging tools, but this modality has yet to be adopted in medical practice. The field is currently led by machine learning models processing acoustical features, which have presented promising diagnostic performances. However, these methods overlook the intricate multi-source nature of audio signals and the underlying mechanisms at play. By addressing this critical gap, the present paper introduces a novel causal framework for validating knee acoustical features. We argue that current machine learning methodologies for acoustical knee diagnosis lack the required assurances and thus cannot be used to classify acoustic features as biomarkers. Our framework establishes a set of essential theoretical guarantees necessary to validate this claim. We apply our methodology to three real-world experiments investigating the effect of researchers' expectations, the experimental protocol and the wearable employed sensor. This investigation reveals latent issues such as underlying shortcut learning and performance inflation. This study is the first independent result reproduction study in the field of acoustical knee health evaluation. We conclude with actionable insights from our findings, offering valuable guidance to navigate these crucial limitations in future research. 
\end{abstract}

\begin{IEEEkeywords}
Knee Acoustic Emissions, Knee Biomarkers, Causal Machine Learning, Shortcut Learning, Explainable-AI, Knee Arthritis
\end{IEEEkeywords}

\IEEEpeerreviewmaketitle

\section{Introduction}
\label{sec:introduction}
Knee health evaluation by examining the vibrations of the joint has been proposed as an alternative to other diagnostic approaches such as medical imaging. This family of methods, also referred to as Vibration Arthrometry, are based on the premise that structural imperfections / damage to an unhealthy knee will generate abnormal vibrations \cite{teague2016novel}. Capturing and characterizing these abnormalities offers the opportunity to develop novel diagnostic tools. In one of the first medical reports on knee acoustic emissions \cite{blodgett1902auscultation}, Blodgett set the tone: Finding patterns useful for diagnosis would prove to be a challenging task. He concluded his report by stretching the main difficulty of the acoustical modality: Unlike other diagnostic tools, such as X-ray imaging, the interpretation of the knee audio signal is significantly based on the \textit{"imagination of the observer"}. 

Blodgett's initial work effectively kicked off efforts to increase the diagnostic value of acoustic signals for knee pathologies. In the 1990s, significant advances in sensor and signal processing technologies led to efforts to gain more insights and understanding of these audio recordings \cite{mccoy1987vibration, zhang1992mathematical, zhang1994adaptive}.

The introduction of Machine Learning (ML) shifted the field's scope from identifying, understanding and modelling audio signals to directly developing \textit{features} relevant to several knee conditions, utilizing ML methods for knowledge discovery. The workflow usually adopted is the following \cite{spain2015biomarkers, teague2016novel, krishnan2000adaptive, whittingslow2020knee, gharehbaghi2021acoustic}~: \begin{enumerate*} [label=(\roman*)]
    \item A vibration or acoustical sensor is placed on the skin surface near the region of the knee.
    \item The sensor records vibration/audio signals while the patient performs specific postular transitions, usually unloaded flexion-extension or loaded sitting-to-stand.
    \item After the recording session has finished, the acquired signals are filtered. Filtering usually involves bandpass filtering \cite{whittingslow2020knee}, or discarding regions of signals that are considered to be significantly affected by external artifacts \cite{gharehbaghi2021acoustic}.
    \item  In the final step, the features are extracted from the recordings and used as input to a ML algorithm to infer the health status of the recorded knee, by classifying it in a range between healthy and pathological with respect to a disease. 
\end{enumerate*}
The acoustic features or the model's output is then treated as a "biomarker" to evaluate the patient's knee health. A plethora of methods have emerged following this workflow, presenting optimistic results based on diagnosis classification accuracies as high as 96\% \cite{abbott2013vibration}. 

Despite the overall positive reports, such approaches have not yet been adopted in medical practice~\cite{katz2021diagnosis}. Furthermore, the approach presented in knee acoustical biomarker reports is contradicted by the current medical standards: acoustical biomarkers are reported to be capable of successfully diagnosing knee pathologies without any additional modalities or clinical information. In contrast, medical guidelines recommend a multi-modal approach involving patient history, clinical and imaging examinations \cite{felson2006osteoarthritis, katz2021diagnosis}. In their 2013 review, Abott et al. \cite{abbott2013vibration} critiqued the stagnation of the Vibration Arthrometry field: the medical community approaches the field with skepticism, and knee acoustic emissions have not yet found their place in diagnostic applications. The authors argued that for the field to move forward, better methodological practices are needed to instill trust.

However, ten years later, the validity of vibration arthrometry as a diagnostic tool is based solely on the promising performance of reported ML models trained on data collected through retrospective studies \cite{talari2020retrospective}. Knowledge of how pathological acoustical signals manifest is fairly limited. Nevertheless, classification performance alone is not enough to guarantee robustness in ML knowledge discovery \cite{ball2023ai, ge2023has}. In the context of biomedical ML, external sources of information can inflate the model's performance \cite{degrave2021ai, arias2024analysis, wallis2022clever} resulting in wrong conclusions. The model then learns shortcuts: unwanted patterns that do not correspond to the pathophysiology of the patients. The solution is to identify the data structures which lead the model to learn these spurious relations and isolate them. However, in vibration arthrometry, we are still far from what has been achieved in other biomedical fields where there is prior medical knowledge of how body structures as well as pathophysiological conditions are manifested in the corresponding modality. For example, in medical imaging, there is a general understanding of bone and soft tissue structures, information introduced by the machine, as well as common artifacts \cite{hendee2003medical}.

Such insights about knee acoustical signals are considerably limited: identifying the source for each component is not trivial. In fact, it is the goal of the entire acoustical-biomarker-discovery exploration. This realization bears two consequences: 
\begin{enumerate*}[label=(\roman*)]
    \item Identifying possible sources of bias in knee-audio datasets is a significantly more challenging task than in other medical modalities, e.g. images.
    \item Consequently, attribution of statistical differences in acoustical features between populations to a pathophysiological source is not a straightforward process. 
\end{enumerate*}

In this work, we contribute to the field of Knee Vibroarthrometry by introducing a novel causal framework as a tool to robustly validate acoustical knee biomarkers. We argue that the current state of the art in Knee Vibroarthrometry has failed to provide satisfying evidence in support of the field's main hypothesis: unhealthy knee joints produce abnormal vibrations. We begin our exploration by formulating a causal description of the task at hand. Through this theoretical investigation we highlight the main pitfalls of the state-of-the-art workflow and demonstrate that features extracted from knee audio recordings cannot serve as health biomarkers. We support our theoretical approach with our findings on real-world data. To the best of our knowledge, our work is the first independent result reproduction study. Finally, we show how this causal exploration allows for the proposal of a set of strict guarantees that should be met in order to reach robust conclusions. 

\section{Theoretical Exploration}
\label{sec:theoretical_exploration}
\begin{figure*}[h]
    \centering
    \includegraphics[width=0.75\textwidth]{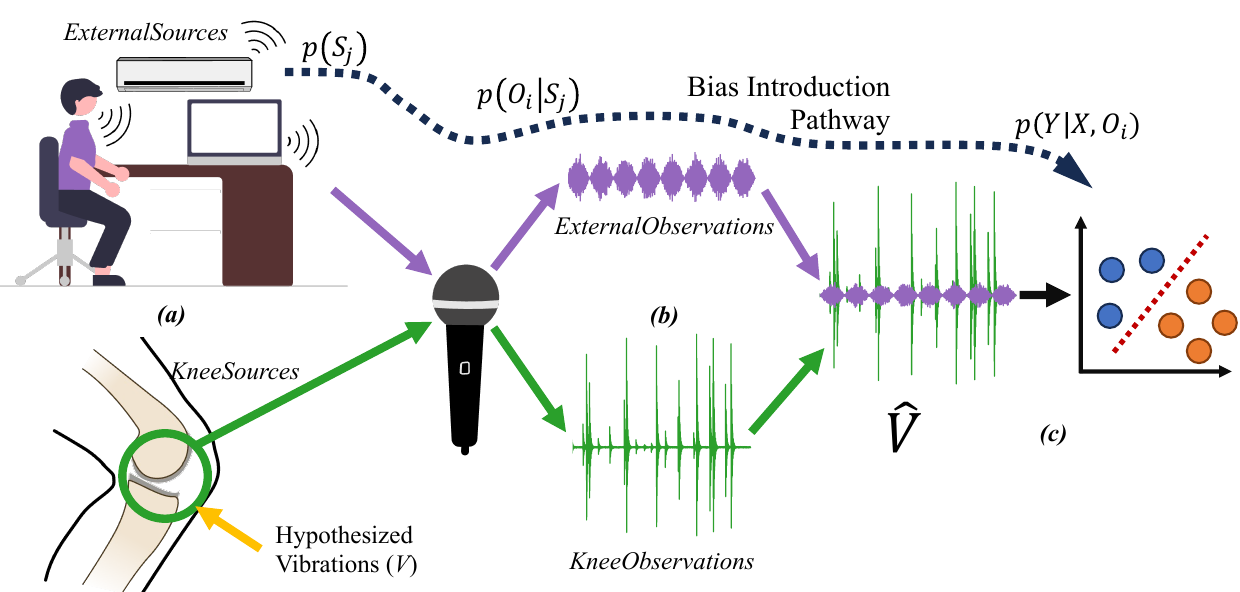}
    \caption{Illustration of our proposed causal framework for investigating and validating acoustic knee biomarkers. \textbf{(a)} Multiple audio source may be present in the experimental environment, including the examined knee. \textbf{(b)} These source will potentially be picked up by the acoustical sensor and may influence the extracted acoustical features. \textbf{c} These external source may introduce bias in the final results, severely boosting the model's performance and leading to wrong conclusions.}
    \label{fig:figure_idea_introduction}
\end{figure*}

We will now attempt to construct a more formalized description of the problem and link this theoretical notation to the workflow adopted by proposed ML methods. An illustrative summary of our approach is presented in Figure \ref{fig:figure_idea_introduction}.

We start with the health condition of the knee, which we will refer to as $H$. Being a complex structure, the joint can be affected by several conditions, such as osteoarthritis, meniscal or ligament tear, etc. \cite{felson2006osteoarthritis}. However, to simplify our analysis, for now we will ignore the vast range of plausible conditions and treat them all the same. Hence, without loss of generality we assume that variable $H$ can take two values: $Healthy$ and $Unhealthy$. A similar approach is also adopted in most publications where unhealthy knees are conditioned by a specific pathology, e.g. meniscus tear \cite{whittingslow2020acoustic}. Furthermore, our approach can also be extended to multi-class classification \cite{galar2011overview}.

The joint health, $H$, will affect the structural integrity of the knee \cite{felson2006osteoarthritis}, and the general consensus is that by extension the vibrations formed in the joint region during motion will also be affected. It is important here to emphasize that this consensus is a hypothesis that needs to be supported by evidence. In an ideal scenario, a vibration sensor should be placed on the affected component of the knee (e.g. meniscus or ligament) to capture these hypothesized vibrations. We refer to these ideal vibration signals as $V$. Thus the main hypothesis can be stated as:

\begin{equation}\label{eq:main_hypothesis}
H \xrightarrow[]{} V
\end{equation}

Unfortunately, such an ideal observation is practically infeasible due to the complexities involved in designing and implementing the experimental setup and doing so without interfering with the observed mechanism. In practice, the vibration sensor is placed non-intrusively on the surface of the skin, in strategic areas near the knee \cite{teague2016novel, mascaro2009exploratory, befrui2018vibroarthrography}. The goal of this experimental setup is to capture the best possible approximation of $V$, $\tilde{V}$. Thus, the Main Hypothesis of Eq. \ref{eq:main_hypothesis}, transforms into the real-world hypothesis:

\begin{equation}\label{eq:main_hypothesis_real_world}
H \xrightarrow[]{} \tilde{V}
\end{equation}

This transformation bears the implicit assumption that $\tilde{V}$ approximates the ideal vibration $V$ in a satisfactory manner. As such, Eq. \ref{eq:main_hypothesis} is equivalent to  Eq. \ref{eq:main_hypothesis_real_world} under the approximation constraint:

\begin{equation}\label{eq:aprox_constraint}
\tilde{V} \approx V
\end{equation}

On the contrary, if this constraint cannot be guaranteed, the equivalence between Eq. \ref{eq:main_hypothesis} and Eq. \ref{eq:main_hypothesis_real_world} cannot be guaranteed either. Of course, robustly validating Eq. \ref{eq:aprox_constraint} presents challenges since access to the actual $V$ is infeasible. Thus, in a real-world setting, we could only require evidence hinting at this constraint. We will elaborate on this issue later on. 

Having acquired the observed signal $\tilde{V}$, the next step is to extract features to describe its qualitative properties that are hypothesized to be linked to the patient's knee health status. An intermediate step of filtering is also usually employed before feature extraction; however, for simplicity let us assume that $\tilde{V}$ has already been filtered and contains the final vibration approximation signal accompanied by all necessary guarantees, i.e. Eq. \ref{eq:aprox_constraint}. Let $g$ be the function that maps the recording $\tilde{V}$ to a set of features $X$, then:

\begin{equation}\label{eq:feature_mapping_function}
X = g(\tilde{V})
\end{equation} In the final step a classifier, $f(\cdot)$ is trained:

\begin{equation}\label{eq:classifier_output}
Y = f(X) = f \circ g (\tilde{V})
\end{equation} so that a loss function between $Y$ and $H$ is minimized. There are many works achieving satisfying levels of classification accuracy \cite{abbott2013vibration}. Interpreting these high classification results is a non-trivial task. The most common argumentation is that they provide evidence supporting the original hypothesis Eq. \ref{eq:main_hypothesis_real_world}, and by extension Eq. \ref{eq:main_hypothesis}. The conclusion is indeed that
\begin{enumerate*}[label=(\roman*)]
    \item health deterioration of the knee can affect the vibrations it emits and 
    \item the experimental setup is capable of capturing and identifying these vibrations.
\end{enumerate*}

Here, we propose a more reserved and pragmatic interpretation. In fact, experiments in small datasets show that $Y$, and by extension (Eq. \ref{eq:classifier_output}) $\tilde{V}$, are correlated with $H$. As such, according to the Common Cause Principle \cite{peters2017elements}, there exists a third variable, $Z$, that causes the dependence between $\tilde{V}$ and $H$ and the conditioning of which renders $\tilde{V}$ and $H$ independent. In the extreme case, $Z$ can be the same as $H$, which is the case adopted by current literature. Making this claim however requires proper guarantees for Eq. \ref{eq:aprox_constraint} in order to reduce the possibility of a third external source of information. That in combination with high $\tilde{V} - H$ correlations could provide strong evidence in support of the argument that $Z$ is indeed $H$ further supporting the original hypothesis from Eq. \ref{eq:main_hypothesis} and \ref{eq:main_hypothesis_real_world}. 

Let's now elaborate on the classification task starting with Eq. \ref{eq:classifier_output} describing the output of the classifier inferring a knee's health by its acoustical observations as input. Observation $X$ is inherently a multi-source signal, and thus exploration of Eq. \ref{eq:classifier_output} should take into consideration $X$'s multi-source nature.

We begin our exploration with the observation $\tilde{V}$, and its representation $X$ in an appropriate domain. A finite set of $N$ sources $Sources = {S_0, ..., S_{N-1}}$ which are observable by our sensor contribute to $\tilde{V}$ (Figure \ref{fig:figure_idea_introduction}a). For each source $S_i$ our sensor observes a corresponding observation $O_i$ forming the finite set of $N$ observations $Observations = {O_0, ..., O_{N - 1}}$. The distinction between the $Sources$ and the $Observations$ is equivalent to that between $V$ and $\tilde{V}$. $Sources$ refer to real-world events, the knowledge of which requires perfect prior information about the world in which the experiments take place. On the contrary, $Observations$, is the set of signal sources that the sensor observes. 

To give more insight on the introduced notation, let us consider the example of the knee acoustical sensor. The event $Source = Knee$ signifies that the knee produces a vibration during the acquisition of the observation $\tilde{V}$. $Observation = Knee$ refers to the event of the sensor capturing the knee and hence $X$ can partly be attributed to the knee. Thus, the probability $p(Observation = Knee \mid Source = Knee)$, describes the sensor's ability to capture a knee acoustic event. Similarly $Observation = Knee \mid Source = \overline{Knee}$ is impossible to happen. Now imagine that in the experimental environment there is an additional external mechanism generating sound waves, for example, the room's ventilation or air-conditioner (A/C) (Figure \ref{fig:figure_idea_introduction}a). Then $p(Observation = \overline{A/C} \mid Source = (Knee, A/C))$ describes the device's ability to filter out the external noise of the A/C. 

For the classification task, given an acoustical observation $\tilde{V}$ we need to estimate the most probable knee health status $Y$:

\begin{equation}\label{eq:classifier_probability}
    p(Y \mid X) = p(Y \mid g(\tilde{V}))
\end{equation}

We consider the multi-source scenario and the notation introduced in Eq. \ref{eq:classifier_probability}. Also, for now the observations $O_i, O_j\quad \allowbreak  \forall O_i, O_j \quad \in \quad Observations$ are considered disjoint and similarly all sources $S_i, S_j \quad \forall S_i, S_j \quad \in \quad Sources$. In the real world this assumption cannot be guaranteed; in the previous example the A/C could operate at the same time as the subject mobilizes the knee joint. We will deal with this scenario later. For now, we can rewrite Eq. \ref{eq:classifier_probability} as:

\begin{equation}\label{eq:classifier_probability_split}
    p(Y \mid X) = \sum_{i} p(Y \mid X, O_i)p(O_i \mid X)
\end{equation}

The first term of the sum, $p(Y \mid X, O_i)$, describes a model trying to infer $Y$ from a sample $X$ attributed to observation $O_i$. The second term, $p(O_i \mid X)$, is essentially a source identification model, attributing the sample $X$ to observation $O_i$. Through Eq. \ref{eq:classifier_probability_split}, the initial source-neglecting model is rewritten into a source-aware model that performs source separation and source-specific classification. Using Bayes' rule, we elaborate on the source identification term, $p(O_i \mid X)$:

\begin{equation}\label{eq:source_identification_probability}
    p(O_i \mid X) = \frac{p(X \mid O_i) \cdot p(O_i)}{p(X)}
\end{equation} 

$p(O_i)$ denotes the probability that observation $i$ is observed in our experimental setting using the sensor. Intuitively, we can consider that capturing an observation from source $S_i$ requires that the actual source $S_i$ occurs, $p(S_i)$, and that our sensor is capable of observing this event when it is happening, $p(O_i | S_i)$. As such, considering all possible sources $S_j$, we can split $p(O_i)$ as:

\begin{equation}\label{eq:source_probability}
    p(O_i) = \sum_{j} p(O_i \mid S_j) p(S_j)
\end{equation}

Putting together Eq. \ref{eq:classifier_probability}, \ref{eq:classifier_probability_split}, \ref{eq:source_identification_probability} and \ref{eq:source_probability}:

\begin{equation}\label{eq:classifier_probability_detailed}
    \begin{split}
        p(Y \mid X) = \frac{1}{p(X)} \sum_{i} p(Y \mid X, O_i)p(X \mid O_i) \\
        \cdot \sum_{j} p(O_i \mid S_j) p(S_j)
    \end{split}
\end{equation}

For the use-case of classifying the health status of a knee, $H$, from an acoustical sample $X$, we can split the set of Sources - Observations into two subsets: those that can be attributed to the knee, and those that are attributed to external phenomena, e.g. A/C :

\begin{equation}\label{eq:classifier_probability_detailed_knee_split}
    \begin{split}
        p(Y \mid X) \propto \sum_{\substack{Knee\\ Observations}} p(Y \mid X, O_i)p(X \mid O_i) \\
        \cdot \sum_{j} p(O_i \mid S_j) p(S_j) \\
        + \sum_{\substack{External\\ Observations}} p(Y \mid X, O_i)p(X \mid O_i) \\
        \cdot \sum_{j} p(O_i \mid S_j) p(S_j)
    \end{split}
\end{equation}

Eq. \ref{eq:classifier_probability_detailed} interestingly reveals that: inferring a knee's health condition by its acoustic observations implicitly requires understanding of the knee behavior and the sensor's ability to capture it. 

Eq. \ref{eq:classifier_probability_detailed_knee_split} shows how a model $f(X)$ inferring $p(Y \mid X)$ can still draw information from external sources as long as $p(Y \mid X, O_i)$ is informative for at least one of them, creating a Bias Introduction Pathway (Figure \ref{fig:figure_idea_introduction}). This equation also reveals a set of strategies to achieve a robust experimental setting. In the Approximation Constraint, Eq. \ref{eq:aprox_constraint}, we required knowledge of the ideal knee observation $V$, which was considered physically infeasible. Here, we require to minimize the effect of the second summation term, so as to ensure that no external source interferes with the classification task, breaking the Bias Introduction Pathway. This can be achieved by optimizing in three different directions:

\begin{enumerate}
    \item $p(Y \mid X, O_i)$: External observations should be random and not correlated with knee health.
    \item $p(O_i \mid S_j)$: The sensor should be shielded from interference sources, so that external signals do not affect knee recording.
    \item $p(S_j)$: Ideally, there should not be any external sources of interference in the experimental setting. The only source of information present in the audio recordings, $\tilde{V}$ should be the knee mechanism.
\end{enumerate}

For each external observation $O_i$ in the world of our experiments, ensuring at least one of these conditions is enough to provide a robust dataset, safe from external biases. Achieving this is not trivial since one needs to first identify all possible sources of external interference and then study their effect on the experiment. However, this task is considerably more feasible than acquiring direct access to the ideal recording $\tilde{V}$.

An important assumption that was necessary to derive Eq. \ref{eq:classifier_probability_detailed} was that the sources in the set of $Sources$ are pairwise disjoint, and similar for the observations. In the real world however this assumption does not necessarily hold true, and multiple events can occur during the recording of an observation $\tilde{V}$. To overcome this limitation, we define $Sources$ as follows. First, we define the finite set of all $N$ sources occurring in the experiment $S$, the complementary set $\overline{S} = \{\overline{x}, x \in S\}$ and their union $A = S \cup \overline{S}$. Now consider one subset $M_i \subset A$ such that:

\begin{equation}
\label{eq:set_cardinality}
    |M_i| = N
\end{equation}
and
\begin{equation}
\label{eq:set_condition_of_elements}
    \forall x_i, x_j \in M_i, x_i \neq \overline{x_j}
\end{equation}

Define the event $S_i$ as the intersection of all elements of $M_i$, $S_i = \bigcap\limits_{x_j \in M_i} x_j$. Due to Eq. \ref{eq:set_cardinality} and \ref{eq:set_condition_of_elements}, given two different sets $S_i$ and $S_j$, they are disjoint. Finally the $Sources$ set is defined as the union of all possible $S_i$. The set of $Observations$ can be defined in a similar way. For example, in the case where we only consider the sources of the Knee and the A/C, instead of defining $Sources = \{Knee, A/C\}$, $Sources$ should be defined as:

\begin{equation}
    \begin{split}
        Sources = \{(Knee, A/C), (\overline{Knee}, A/C), \\
                    (Knee, \overline{A/C}), (\overline{Knee}, \overline{A/C})\}
    \end{split}
\end{equation}

\section{Real-world Lessons}
\label{sec:real_life_lessons}
So far, we have established a framework to describe and characterize the task at hand and the tools employed to tackle it, i.e. classification. However, our analysis has remained mostly theoretical. In this section we complement this exploration with three real-life study cases. Simultaneously we provide key takeaways from our investigation and literature regarding best practices when designing knee acoustic emission experiments and evaluating their results.

\subsection{The Relevance of the Expectations}
\label{sec:a_semi_realistic_example}
\begin{figure}[h]
    \includegraphics[width=0.49\textwidth]{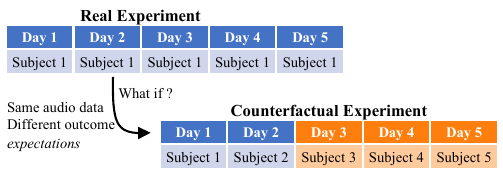}
    \caption{Illustration of our counterfactual experiment. By changing the expectation on the output (asking what if) the same audio data are interpreted differently. This interpretation is not necessarily causally linked to the underlying knee mechanism we are trying to describe.}
    \label{fig:counterfactual_experiment_illustration}
\end{figure}

We begin with a counterfactual thought experiment involving the design of a study to identify and evaluate acoustical biomarkers. The question we raise is: Will acoustical features reveal knee structural differences, even when they should not? We show how the researchers' expectations regarding the experiment's output, i.e. how the samples are split in health/unhealthy groups, can influence the interpretation of the features. An illustration of the counterfactual thought-experiment is presented in Figure \ref{fig:counterfactual_experiment_illustration}.

The exploration is set up as follows. We perform a real experiment with a single subject who underwent measurements on the right knee for five consecutive days. Following a similar workflow as the ones proposed in the relevant literature, we employed a wearable acoustical device to record knee audio signals. The wearable device uses two microphones to record signals from the sub-patella region: one on the medial and one on the lateral side of the knee. On each day the subject performed 6 unloaded reciprocal knee flexion-extensions, an exercise that is usually used for this purpose \cite{teague2016novel, gharehbaghi2021acoustic, whittingslow2020knee, whittingslow2020acoustic, zhang1994adaptive, befrui2018vibroarthrography, jeong2018b, hersek2017acoustical, ozmen2021novel}. After collection, a set of audio features was extracted, following the work of \cite{gharehbaghi2021acoustic, whittingslow2020knee, befrui2018vibroarthrography, hersek2017acoustical, ozmen2021novel}.  All recordings were taken approximately at the same time of the day, between 12:00 and 12:30 and no injury was reported throughout the week. Hence, we have no reason to believe that there was any mechanical degradation in the subject's knee mechanism that would affect its acoustic emissions throughout the course of the experiment. Under these assumptions, any variance in the audio features is interpreted not as fluctuations in the knee mechanism but rather as variance in the measurement process.

\begin{figure}[h]
    \centering
    \includegraphics[width=0.5\textwidth]{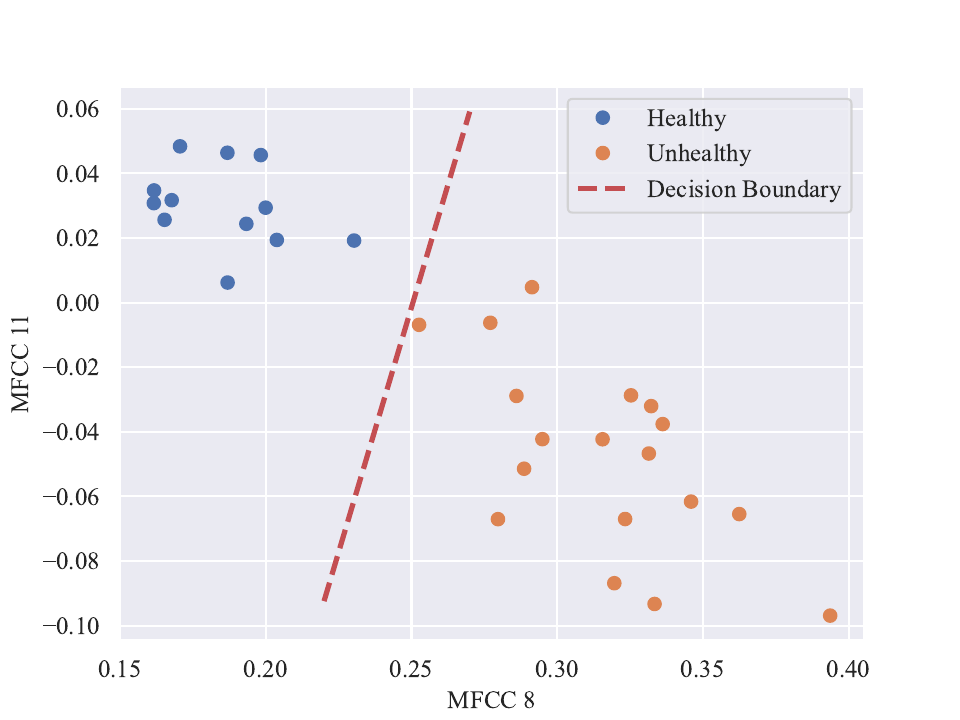}
    \caption{Separation of Healthy / Unhealthy subsets based on MFCC8 and MFCC11.}
    \label{fig:MFCC_8_vs_MFCC_11}
\end{figure}

Using the same audio recordings, we ask a counterfactual question, changing the researchers' expectations: What if the data had been recorded from 5 different subjects? In this scenario, the first 2 knees are healthy, while the next 3 are unhealthy. Now we are looking to identify the most relevant audio features that will serve as our proposed knee health biomarkers. After our analysis, we conclude that the Mel Frequency Cepstral Coefficient (MFCC) features \cite{abdul2022mel}, specifically MFCC 8 and MFCC 11, have a good discriminant validity, as they show a clear separation between the two populations (Figure \ref{fig:MFCC_8_vs_MFCC_11}). Furthermore, similar audio features have already been successful in discriminating healthy and unhealthy subjects \cite{gharehbaghi2021acoustic}. The study of this counterfactual thought experiment is concluded with very positive remarks. We have managed to identify promising audio biomarkers, capable of identifying unhealthy knees with a Leave-One-Subject-Out (LOSO) Cross Validation accuracy of 96\%. 

In this experiment, by construction, we know that the audio features do not characterize biomechanical degradation. Yet, in the counterfactual case, the interpretation of the high classification accuracy led to (falsely) attributing the feature information to the knee mechanism. 
\\
\\
\textbf{Takeaway: High classification accuracy alone is not enough to establish knee health biomarkers.}
\\

Conclusions such as the one made in the counterfactual scenario fail to take into consideration the multi-source nature of the audio recordings: all the information contained in the signal has been attributed to the observed knees and their health status. However, referring back to Eq. \ref{eq:classifier_probability_detailed_knee_split}, there are two terms that could potentially contribute to this classification: one whose source is the joint mechanism and another one which should be attributed to external sources, w.r.t. the joint sources. 

Proper source attribution in Eq. \ref{eq:classifier_probability_detailed_knee_split} is crucial: by attributing all information to the knee we have managed to identify structural differences where we should not have. Thus, a causal exploration of the phenomena that affect the audio features identified as relevant biomarkers should not be considered as future work \cite{befrui2018vibroarthrography, whittingslow2020knee}, but rather a necessary step in the researchers' exploration. 
\\
\\
\textbf{Takeaway: Causal investigation on the audio features is imperative to claim them as biomarkers.}
\\
\\
Such an investigation is not necessarily a trivial task. In the next two study-cases we present two examples of acoustical features causal study.

As a final note, two points need to be addressed. The first one is the population size which is admittedly small (5 subjects). Nevertheless, similar knee acoustic emission studies with comparable sample sizes exist in the literature \cite{khan2016acoustic, whittingslow2020acoustic, shark2022discovering, khokhlova2021motion, jeong2018b, khokhlova2022test, ozmen2021novel}. Furthermore, a larger population size could help address the issue in this particular case: it decreases the probability that $p(Y \mid X, O_i)$ is informative for external sources. However, it does not necessarily guarantee lack of bias in the data. As we demonstrate in the next section a larger population size can still be affected by external sources.
\\
\\
\textbf{Takeaway: Larger sample size helps, but cannot substitute proper causal investigation. }
\\
\\
The second point is that the audio data remained the same in the counterfactual exploration with only the output - expectations changing. The counterargument here is that in the scenario with multiple subjects, Healthy and Unhealthy, other components related to the health variable would have been added in the audio features. This is also what the state of the art usually adopts as a direction. However it is crucial here to stress that this is just a hypothesis which needs to be validated. It should not be considered as prior information. What we have shown in this example is that this hypothesis is not necessary for the health classification accuracy to be high.

\subsection{The Relevance of the Protocols}
\label{sec:a_real_life_cautionary_tale}

This case-study is a real-world case of a publicly available dataset \cite{whittingslow2020knee}. 43 subjects were recruited for the study: 18 healthy and 25 with Juvenile Idiopathic Arthritis (JIA) \cite{ording2015joints}. Five of the 25 JIA subjects were not included in the public dataset. We replicate the workflow reported in the original publication:
\begin{enumerate}
    \item Bandpass filter the audio signal in the range 250 Hz - 10 kHz.
    \item Extract the most significant features, as reported in \cite{whittingslow2020knee}.
    \item Train and validate a linear regression model using the LOSO strategy. 
\end{enumerate}

We achieved an average exercise-repetition accuracy of 51.28\%, a significant drop from the reported 80.6\%. To investigate further, we performed the following explorations:
\begin{enumerate*}[label=(\roman*)]
    \item manual inspection of the audio signals and
    \item per frequency-range classification. 
\end{enumerate*}

\begin{figure*}[h]
     \centering
     \begin{subfigure}[b]{0.32\textwidth}
         \centering
         \includegraphics[width=\textwidth]{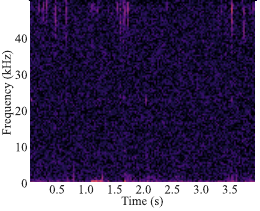}
         \caption{}
         \label{fig:inan_example_healthy_audio_spectrogram}
     \end{subfigure}
     \hspace{1mm}
     \begin{subfigure}[b]{0.32\textwidth}
         \centering
         \includegraphics[width=\textwidth]{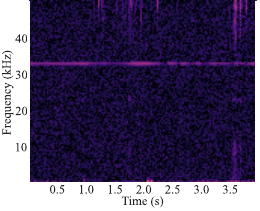}
         \caption{}
         \label{fig:inan_example_unhealthy_audio_spectrogram}
     \end{subfigure}
     \hspace{1mm}
     \begin{subfigure}[b]{0.32\textwidth}
         \centering
         \includegraphics[width=\textwidth]{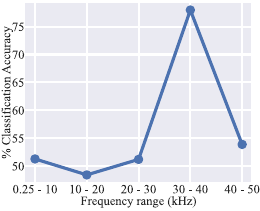}
         \caption{}
         \label{fig:accuracy_vs_frequency_range_20most_relevant_features}
     \end{subfigure}
 
    \caption{External 33kHz interference and its effect on the health classification task. Representative time-frequency representations of the audio recordings of a Healthy \textbf{(a)} and an Unhealthy \textbf{(b)} individual. The constant component interference characterizing the unhealthy samples is visible at around 33kHz. \textbf{(c)} The accuracy of the knee health classification as a function of the frequency range demonstrating the effect of external interference on the classification task.}
    \label{fig:inan_example_healthy_vs_unhealthy}
\end{figure*}

\textbf{Manual Inspection}. We performed an audio-visual inspection of the recorded signals in the time-frequency domain. A constant frequency component centered around 33 kHz with a high prevalence was detected in the unhealthy population, a representative example is presented in Figure \ref{fig:inan_example_unhealthy_audio_spectrogram}. However, this component was not present in any of the healthy samples (Figure \ref{fig:inan_example_healthy_audio_spectrogram}), with the exception of one healthy subject. This signal is constant throughout the duration of the recording, even when the legs are inactive. Thus, we consider impossible the scenario that it is generated by the joint's internal mechanism. Rather, there has to be an external interference. 
\\
\\
\textbf{Takeaway: Thorough manual audio-visual inspection helps in the causal attribution investigation.}
\\

\textbf{Per Frequency-Range Classification}. We elaborate on the previous findings by determining the frequency range that is more informative for the knee's health status, i.e. it achieves higher classification accuracy. We split the available range 250 Hz - 50 kHz of the signal into 5 groups of 10 kHz. For each frequency range group, we follow the same procedure as in \cite{whittingslow2020knee}: features are extracted for the corresponding frequency range, and a linear model is trained and validated using LOSO cross validation. Figure \ref{fig:accuracy_vs_frequency_range_20most_relevant_features} presents the results: classification accuracy as a function of the frequency components in the inputs of the classifier. Average accuracy peaks in the 30 kHz - 40 kHz range at 77.9\%, while for the rest of the groups it remains close to 50\%. 

As discussed in Section \ref{sec:theoretical_exploration}, the Common Cause Principal is crucial in interpreting our results. The external constant interference, $Z$, causes the dependence between the classifier's output $g(\tilde{V})$ and the knee health status $H$. Since the knee cannot produce such an interference, $Z$ is not the same as $H$. By filtering out the 33kHz component, effectively conditioning on $Z$, $\tilde{V}$ and $H$ become independent. It is also important to note that the constraints resulting from Eq. \ref{eq:classifier_probability_detailed_knee_split}, as discussed in Section \ref{sec:theoretical_exploration}, are also violated. We can now summarize the Bias Introduction Pathway proposing solutions to breaking it. For the second sum of Eq. \ref{eq:classifier_probability_detailed_knee_split} describing the effect of external sources :
\begin{enumerate}
    \item $\boldsymbol{p(S_j)}$\textbf{:} There is obviously an external interference. \\
    \textbf{Potential Solution:} Remove the external interference from the experiment environment. 
    \item $\boldsymbol{p(O_i \mid S_j)}$\textbf{:} The acoustic sensor observes this external interference. \\
    \textbf{Potential Solution:} Isolate the sensor and electronics from the detected interference. Depending on the nature of the external source filtering can also be performed post-hoc through digital filtering, if it is not affecting the actual knee components.
    \item $\boldsymbol{p(Y \mid X, O_i)}$\textbf{:} The external source is not random and provides information to the classifier about the health status of the knee. \\
    \textbf{Potential Solution:} Randomizing the observation of the external component w.r.t. the health status of the knee would prevent bias, even if external interference was present. For example, if it was present in half of the healthy and unhealthy populations and not present in the rest of the subjects. 
\end{enumerate}
\textbf{Takeaway: The same environmental conditions should be guaranteed for the entire trial population.}
\\
\\
Such inconsistencies may arise either from ambient noises, e.g. machinery present in the room, or from electronic component noise from within the signal acquisition setup, e.g. power source. 

Regardless the solution, the first and most difficult step in tackling the bias issue is detecting and identifying it. In this case-study, manual inspection of the signals, using intuitive time-frequency representations was crucial in detecting the 33 kHz constant component. Equally important is observing its behavior and evaluating the possibility that it is generated by the knee mechanism. Such insights require prior knowledge on the experimental setting and on the biomechanical properties of the knee joint. It is difficult or even impossible to acquire this level of information just from the trained model alone.
\\
\\
\textbf{Takeaway: Knowledge of the experimental conditions is necessary to attribute signal components to suspected sources.}
\\

\subsection{The Relevance of the Sensors}

In the third and final case study, we use a custom-made apparatus similar to a knee orthosis to collect acoustic information from the subpatellar medial compartment of the knee joint without hindering the movements of the patient. The apparatus consisted of a silicone-covered frame with embedded electronics to collect and store data from a non-contact acoustic sensor positioned at a constant distance from the area of interest. Institutional review board approval was obtained (BASEC-Nr.: 2020-01031) and all patients signed an informed consent prior to data acquisition. The population consists of 16 subjects undergoing for unilateral total knee replacement surgery: for half of the population, the left leg is the one to be operated, while for the rest half, the right knee. For the sake of consistency with the terminology used in the rest of the manuscript, we will label as \textit{healthy} the non-surgical knees, and \textit{unhealthy} surgical ones. However, it is important to note that we do not have specific information about the health status of the not-to-be-operated legs, and our objective is reduced to distinguish between groups. Patients performed eight unloaded and consecutive flexion-extension movements of the knee at a self-selected comfortable pace while wearing an apparatus on each leg, to record knee sounds. One device was always worn on the left knees, which we will refer to as the left device, while another one on the right ones, the right device. The classification task is thus defined as classifying the legs as to-be or not-to-be operated, depending on their acoustic emissions. 

We performed a feature-based analysis similar to the two previous experiments. The acoustical signals were bandpassed in the frequency range 900Hz - 3kHz and then the acoustical features were extracted. We have observed that the family of MFCC features is one of the most relevant ones, so we designed a linear model to classify each knee as surgical or non-surgical based on the MFCC feature of its audio signals. With LOSO cross-validation we achieved a 75\% classification accuracy. Neglecting Eq. \ref{eq:aprox_constraint} and \ref{eq:classifier_probability_detailed_knee_split} we can conclude that acoustic emissions can predict the knee undergoing surgery. 

Just like in Sections \ref{sec:a_semi_realistic_example} and \ref{sec:a_real_life_cautionary_tale}, we introduce our causal approach. First, we investigate whether the two devices themselves introduced bias. To this end we pose two questions:
\begin{enumerate}
    \item Can we infer the device from the same audio features?
    \item Does the knee health status classification performance change when conditioning on the device?
\end{enumerate} Then, we explore the mechanism - structures in the data allowing the separation between healthy and unhealthy knees.

To answer the first question, we train a classifier to infer the device based on the audio features. LOSO validation results in a device classification accuracy of 87.5\%. Indeed, the same audio features can differentiate between left and right leg devices. 

\begin{figure}[h]
    \includegraphics[width=0.5\textwidth]{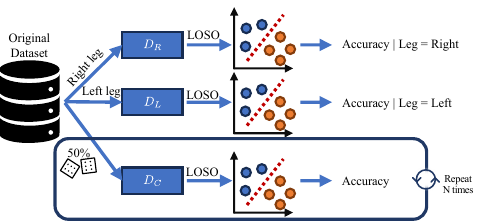}
    \caption{Illustration of the process to assess the effect of the device on the health classification performance by conditioning on the device.}
    \label{fig:experiment_3_illustration}
\end{figure}

For the second question we condition on the device side (Figure \ref{fig:experiment_3_illustration}) forming the following subsets from the original dataset:
\begin{enumerate}
    \item One set consisting only of the left legs, $D_L$,
    \item one only of the right ones, $D_R$ and
    \item a third sample as control, $D_C$, to account for the effect of sample size, since the sample size in each subset, $D_L, D_R$, is approximately halved compared to the original dataset. $D_C$ is formed by randomly sampling 50\% of the samples in the original dataset.
\end{enumerate}

We train and validate (LOSO) a linear classifier from scratch for each of the three subsets: $D_L, D_R, D_C$. Since $D_C$ is a random sampling of the original dataset, we repeat the experiment for $D_C$ $10.000$ times to account for variation introduced between different random splits. A different linear classifier is trained on each iteration.

\begin{figure}[h]
     \centering
     \begin{subfigure}[b]{0.4\textwidth}
         \centering
         \includegraphics[width=\textwidth]{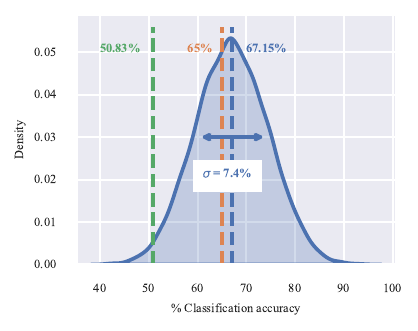}
         \caption{}
         \label{fig:50percent_random_legs_out}
     \end{subfigure}
     \begin{subfigure}[b]{0.4\textwidth}
         \centering
         \includegraphics[width=\textwidth]{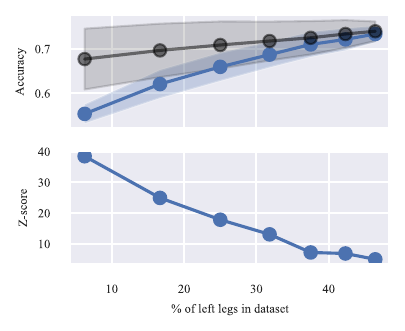}
         \caption{}
         \label{fig:accurcy_dependence_on_left}
     \end{subfigure}
 
    \caption{Dependence of the classification accuracy on the device. \textbf{Top:} Conditioning on the right device the accuracy drops to 50\% (green vertical line), a significant reduction that cannot be explained by the reduced sample size (blue distribution). Conditioning on the left device is presented with an orange vertical line. \textbf{Bottom:} Classification accuracy is dependent on the percentage of sample of the left leg in the dataset (blue). This dependence cannot be explained by the reduced amount of samples (black).}
    \label{fig:inan_example_healthy_vs_unhealthy}
\end{figure}

The results are presented in Figure \ref{fig:50percent_random_legs_out}. We can observe that the effect of subsampling on the accuracy (blue distribution $D_C$, $67.15\% \pm 7.4\%$) does not explain the significant drop to 50\% for $D_R$ (green line).

The dependence of the accuracy on the existence of samples from the left leg is also shown in Figure \ref{fig:accurcy_dependence_on_left}. Here, we begin with the subset $D_R$ and incrementally add $1, 2, ...$ left leg samples. The left legs to be added are randomly selected and the experiment is repeated 500 times. As a point of reference, we also evaluate the classification accuracy if all legs samples had been drawn randomly but with the same total number of legs. Observe how a large percentage (higher than 40\%) of samples from the left leg is necessary for the two distributions to start to converge. 

\begin{figure*}[h]
    \centering
    \includegraphics[width=0.85\textwidth]{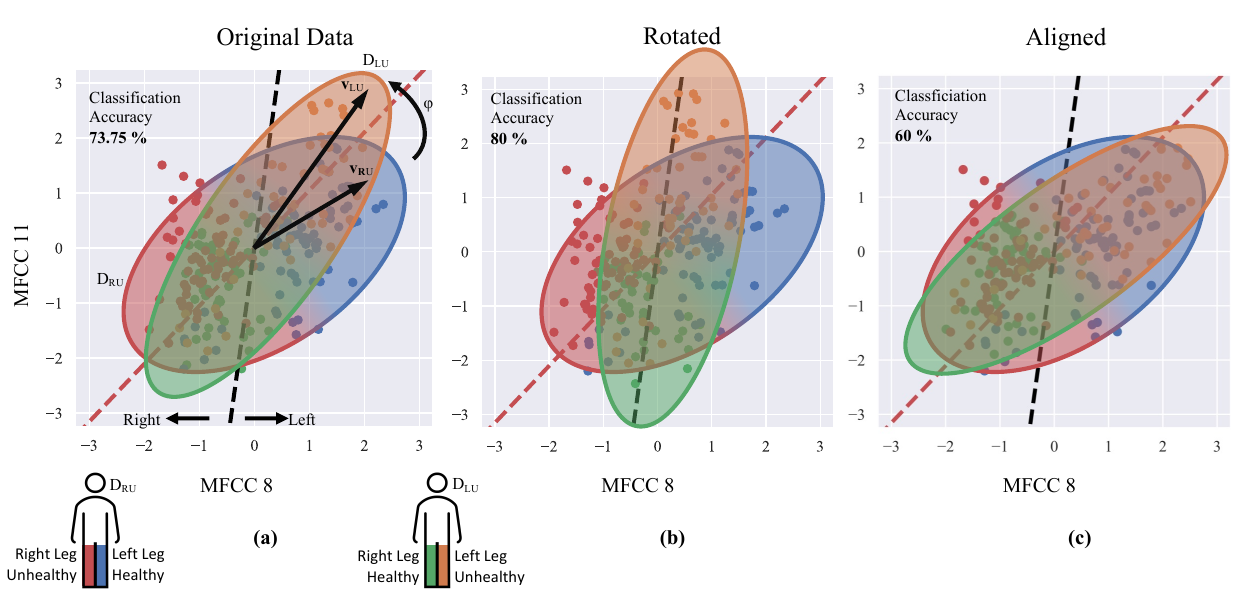}
    \caption{Normalized MFCC8 - MFCC11 feature map. The colored ellipses visualize the two populations $D_{LU}$ and $D_{RU}$. \textbf{(a)} The original data. \textbf{(b)} The samples of $D_{LU}$ have been rotated counter-clockwise to accentuate the separation between the healthy and unhealthy samples. \textbf{(c)} $D_{LU}$ has been rotated clockwise so that the first principal components of $D_{LU}$ ($\textbf{v}_{LU}$) and $D_{RU}$ ($\textbf{v}_{RU}$) align.}
    \label{fig:data_rotation}
\end{figure*}

We will now explore how the three structures of the data allow for the classification between healthy and unhealthy knees. For visualization purposes, we will only focus on two of the most informative MFCC features: MFCC8 and MFCC11 (Figure \ref{fig:data_rotation}) which can achieve 73.75\% classification accuracy. 

\textbf{Structure 1: Population}
The first structure is our population of 16 patients which is divided into two sub-groups:
\begin{enumerate}
    \item 6 patients with left knee unhealthy and right knee healthy, $D_{LU}$ and
    \item 10 patients with left knee healthy and right knee unhealthy, $D_{RU}$.
\end{enumerate}
There are no patients with both legs healthy or both legs unhealthy, breaking the independence between the samples: let $H_{L_i}$ be the health status of the left leg of patient $i$ and $H_{R_i}$ the health condition of their right leg, then $p(H_{L_i} = Unhealthy) = 1 - p(H_{R_i} = Unhealthy)$.

\begin{figure}[h]
     \centering
     \begin{subfigure}[t]{0.48\textwidth}
         \centering
         \includegraphics[width=\textwidth]{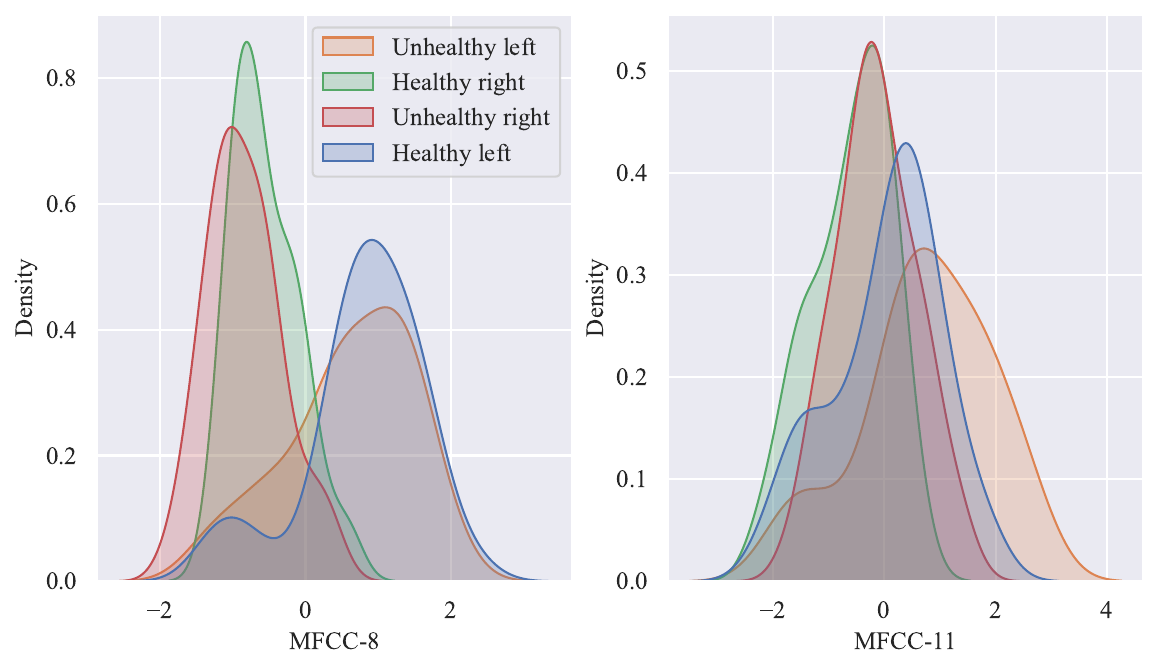}
         \caption{}
         \label{fig:leg_separation_mfcc8_mfcc11}
     \end{subfigure}
     \hspace{1mm}
     \begin{subfigure}[t]{0.48\textwidth}
         \centering
         \includegraphics[width=\textwidth]{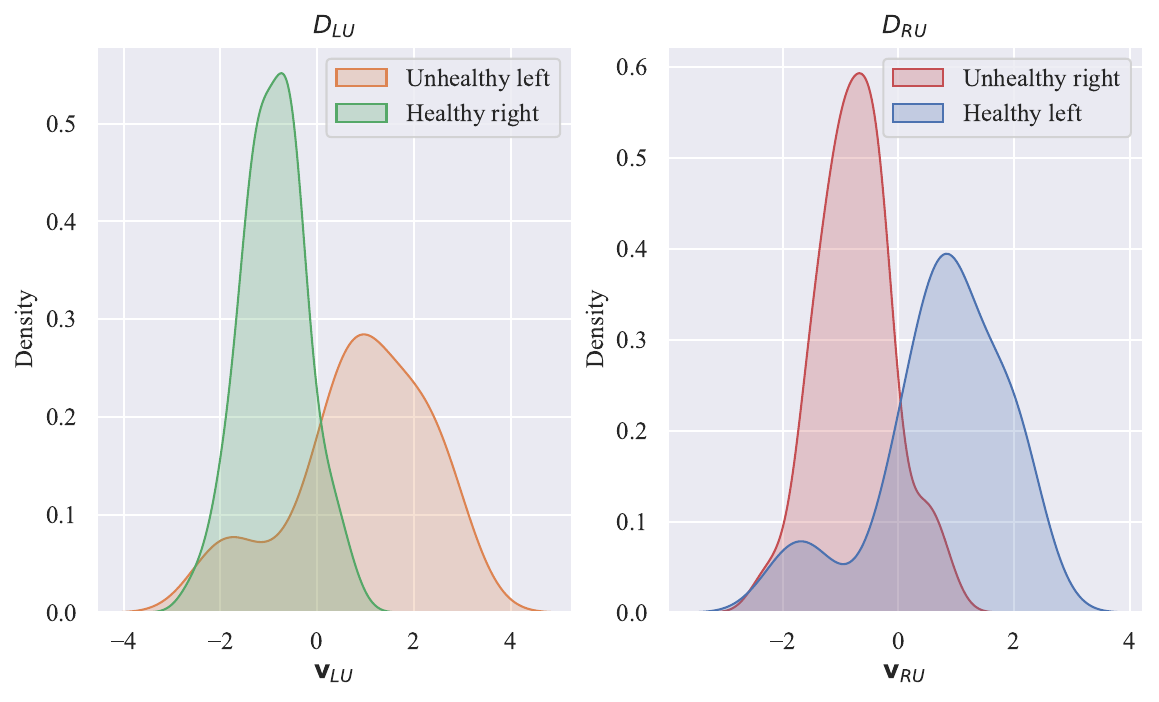}
         \caption{}
         \label{fig:leg_separation_along_main_components}
     \end{subfigure}
 
    \caption{Feature plots indicating separation between left and right leg. \textbf{Left:} Leg separation in the MFCC8 and MFCC11. The features do not differentiate between healthy and unhealthy legs: Healthy Right coincides with Unhealthy Right. \textbf{Right:} Distributions of features when projected along the first main principal component. Leg side separation is evident while health status is independent. For example, similarly to MFCC8 and MFCC11, Healthy Right overlaps with Unhealthy Right and Healthy Left overlaps with Unhealthy left.}
    \label{fig:leg_separation_features}
\end{figure}

\textbf{Structure 2: Device Identification}
The second data structure is the separation between left and right knees. MFCC 8 differentiates between the left and the right device (Figure \ref{fig:leg_separation_features}), MFCC 11 also presents the same separation but the effect is not so strong. As a result, two samples $x_i \in D_{L}, y_i \in D_{R}$ from the same patient, either $D_{RU}$ or $D_{LU}$, tend to be positioned diametrically opposite in the MFCC8-MFCC11 feature map (Figure \ref{fig:data_rotation}a). In fact, for both subsets the first principal component ($\textbf{v}_{RU}, \textbf{v}_{LU}$) distinguishes between the left and right devices, Figure \ref{fig:leg_separation_along_main_components}, while it is independent of the leg's health condition. 

\begin{figure}[h]
    \centering
    \includegraphics[width=0.35\textwidth]{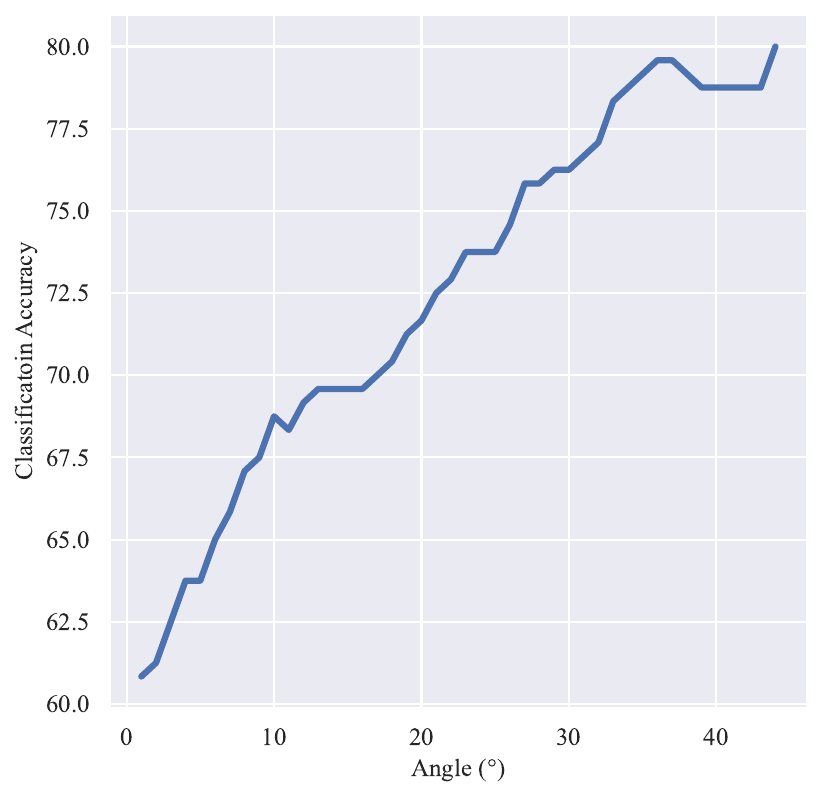}
    \caption{The ability to distinguish between healthy/unhealthy legs is dependent on the distribution shift between the two sub-populations $D_{LU}$ and $D_{RU}$. The classification accuracy is sensitive to the shift and just a few degrees of relative angle can inflate it.}
    \label{fig:aligned_dists}
\end{figure}

\textbf{Structure 3: Distribution shift}
The third and final data structure is the distribution shift between $D_{LU}$ and $D_{RU}$. There is a rotation between $\textbf{v}_{RU}$ and $\textbf{v}_{LU}$ (Figure \ref{fig:data_rotation}a): $\phi(\textbf{v}_{RU}, \textbf{v}_{LU}) = cos^{-1}\left(\frac{\textbf{v}_{RU} \cdot \textbf{v}_{LU}}{|\textbf{v}_{RU}| \cdot |\textbf{v}_{LU}|}\right)$. It is this misalignment, $\phi(\textbf{v}_{RU}, \textbf{v}_{LU})$ which allows the classification between healthy and unhealthy legs. Accentuating the shift between the two sub-populations, Figure \ref{fig:data_rotation}b, results in a better separation and higher classification accuracy (80\%). In contrast, aligning the two populations, $\phi = 0$, results in the inability to distinguish between healthy and unhealthy legs (Figure \ref{fig:data_rotation}c). Interestingly, only a few degrees of distribution shift, $\phi \leq 10^{\circ}$, can result in a significant inflation of the classification accuracy (Figure \ref{fig:aligned_dists}).

Interpreting the distribution shift, $\phi$, is non-trivial. If the effect of knee health on the audio features is considered as prior knowledge, and given the underlying Structures 1 and 2, then $\phi$ is the mechanism through which the healthy and unhealthy knees are separated. Still, this interpretation does not explain the lack of health status separation in the Right device, in contrast to the Left device: the effect of knee degradation should be independent of the device. In any case, in this scenario the results cannot be used as evidence in favor of the Main Hypothesis (Eq. \ref{eq:main_hypothesis}), since it has been used as prior information. If the Main Hypothesis is not used as prior information, but rather as a hypothesis we test against, then there's a second possible interpretation of the results, i.e. $\phi$ is the result of variance between the two subpopulations. To reach a conclusive result further investigation would be needed. Just like in previous examples we can refer to Eq. \ref{eq:classifier_probability_detailed_knee_split} for possible directions:
\begin{enumerate}
    \item $\boldsymbol{p(S_j)}$\textbf{:} There is an interference generated by the device itself. \\
    \textbf{Potential Solution:} Study acoustically the two devices to understand its nature and possibly normalize between the two devices in the physical world.
    \item $\boldsymbol{p(O_i \mid S_j)}$\textbf{:} This device-specific interference is being picked up by the microphones. \\
    \textbf{Potential Solution:} Understanding how it manifests in audio recordings could potentially help us filter it digitally. 
    \item $\boldsymbol{p(Y \mid X, O_i)}$\textbf{:} External interference along with the data structures in our dataset allow for the significant inflation of the classification accuracy. \\
    \textbf{Potential Solution:} We propose three possible solutions:
    \begin{itemize}
        \item Include patients with both legs healthy and both legs unhealthy
        \item Expand study to include a larger population
        \item Randomly shuffle the left and right devices between subjects
    \end{itemize}
\end{enumerate}

Our results indicate that the device indeed functioned as a bias source, inflating the performance of the knee health classification. Unlike the previous example, here the external interference observation was indirect, we studied the data structures in the dataset. Still, our extensive feature-based causal investigation was necessary since high accuracy alone was not enough. We demonstrated how data variances, reminiscent of the first example, can explain the separation between healthy and unhealthy samples. The sample size was also limited, sub-sampling half the population results in only 8 subjects, a sample size similar to the 5 subjects of the first example (Section \ref{sec:a_semi_realistic_example}). 

Unfortunately, we were unable to directly detect the audio component differentiating the devices despite thorough manual audio-visual inspection, in a similar fashion to the 33kHz component of Section \ref{sec:a_real_life_cautionary_tale}. We were also unable to manually identify audio events or patterns that discriminate between healthy and unhealthy knees. However, we managed to manually distinguish \textit{click} sounds which could potentially be attributed to the knee mechanism. 

Similarly to Section \ref{sec:a_real_life_cautionary_tale} all three conditions of Eq. \ref{eq:classifier_probability_detailed_knee_split} were violated. In the experiment there was external interference ($p(S_j)$) which was captured by our experimental device ($p(O_i \mid S_j)$) and introduced bias into the classification task ($p(Y \mid X, O_i)$). We suspect differences between the left- and right-leg devices; the two devices were prototyped manually which could have introduced dissimilarities between them. For example, the physical interface between the microphone and the skin could present slight differences resulting in variations in the frictions on the skin. We can now extend our conclusion from Section \ref{sec:a_real_life_cautionary_tale}: external interference can be generated by both the environment and the recording device itself. 
\\
\\
\textbf{Takeaway: Condition on device-specific audio properties.}

\section{Conclusions}

In this work, we have introduced a comprehensive causal framework to the field of knee vibroarthrometry. Our tools allow us to perform a holistic investigation taking into account various possible sources of information, i.e. the knee mechanism as well as external interference, expanding the current state-of-the-art, which considers only the knee mechanism as the information source. 

We have applied our proposed methodology to three case studies. Our findings challenge the assumption that achieving high classification is sufficient evidence for acoustic knee biomarkers. Across the three experiments, we identified external sources of information that significantly inflated knee health classification performance, leading to a gross overestimation of the ability of the proposed biomarkers to diagnose knee degradation. In two out of three experiments, the proposed biomarkers did not characterize the knee mechanism. In the third one the experimental process was not enough to reach a conclusive interpretation of the result, while external biases were present. 

These outcomes underscore the need for a comprehensive causal investigation of audio features to fully claim them as biomarkers. Meticulous audio-visual inspection of the audio recordings plays an important role in our proposed methodology, unveiling external interference and helping attribute observations to sources, as evidenced in the second example. Complementary to manual inspection, a thorough examination of underlying data structures provides insights into the possible external, wr.t. the knee, confounders. 

Finally, we identified two categories of external sources that can contribute to the health classification task: environmental and device-specific. Therefore, maintaining consistent environmental conditions across the entire population is paramount. Similarly, a thorough examination of the wearable audio acquisition device is essential to ensure that it does not affect the classification task or, at least, its effect is homogeneous throughout the entire data set.

A remaining challenge to be considered is the labeling of the conditions to enhance the clinical relevance of any study. It is of utmost importance to define labels based on actual medical information (e.g., examination, imaging...) while considering that most clinical scoring is based on symptoms, which have a non-linear and non-continuous relation with disease severity. Thus, different approaches should be explored to assess the impact of specific symptoms on the acoustic emissions, and also the influence of the model properties in the label space. For example, a model assuming a smooth output or a sharp threshold between classes may not be appropriate for representing the clinical definition of the target labels.

\bibliographystyle{IEEEtran}
\bibliography{refs.bib}

\end{document}